\documentclass[10pt,conference]{IEEEtran}
\usepackage{amsmath,amssymb,amsthm,bm,bbm,color,cite, comment}

\newenvironment{Proof}[1]{\medskip\par\noindent{\bf Proof:\,}\,#1}{{\mbox{\,$\blacksquare$}\par}}

\newtheorem{theorem}{Theorem}

\newtheorem{lemma}{Lemma}

\newcommand{\ua}{\underline{a}}
\newcommand{\ub}{\underline{b}}
\newcommand{\uA}{\underline{A}}
\newcommand{\ucA}{\underline{\mathcal{A}}}

\newcommand{\COmega}{\Omega^{(\theta)}_{p^{(n)}} (X^nY^nZ^n|M_n)}
\newcommand{\COmegaPtoP}{\Omega^{(\theta)}_{p^{(n)}} (X^nY^n|M_n)}
\newcommand{\BComega}{\overline{\Omega}_n^{(\theta)}(W)}

\newcommand{\cC}{\mathcal{C}}

\newcommand{\cM}{\mathcal{M}}
\newcommand{\cP}{\mathcal{P}}

\newcommand{\cU}{\mathcal{U}}
\newcommand{\cX}{\mathcal{X}}
\newcommand{\cY}{\mathcal{Y}}
\newcommand{\cZ}{\mathcal{Z}}
\newcommand{\mP}{\mathrm{P}}
\newcommand{\mPr}{\mathrm{Pr}}
\newcommand{\mE}{\mathrm{E}}

\IEEEoverridecommandlockouts
\allowdisplaybreaks

\begin{document}

\title{Partial Strong Converse for the Non-Degraded Wiretap Channel\thanks{This work was supported by NSF Grants CNS 13-14733, CCF 14-22111, CCF 14-22129, and CNS 15-26608.}}

\author{\IEEEauthorblockN{Yi-Peng Wei \qquad Sennur Ulukus}
\IEEEauthorblockA{Department of Electrical and Computer Engineering\\
University of Maryland, College Park, MD 20742\\
{\it ypwei@umd.edu \qquad ulukus@umd.edu} }}

\maketitle

\begin{abstract}
We prove the partial strong converse property for the discrete memoryless \emph{non-degraded} wiretap channel, for which we require the leakage to the eavesdropper to vanish but allow an asymptotic error probability $\epsilon \in [0,1)$ to the legitimate receiver. We show that when the transmission rate is above the secrecy capacity, the probability of correct decoding at the legitimate receiver decays to zero exponentially. Therefore, the maximum transmission rate is the same for $\epsilon \in [0,1)$, and the partial strong converse property holds. Our work is inspired by a recently developed technique based on information spectrum method and Chernoff-Cramer bound for evaluating the exponent of the probability of correct decoding.
\end{abstract}

\section{Introduction}
We consider the discrete memoryless \textit{non-degraded} wiretap channel, in which a transmitter wishes to send messages to a legitimate receiver while keeping the messages secret from an eavesdropper. The wiretap channel was first studied in \cite{Wyner_WTC_75} with the assumption that the wiretap channel is degraded, and the secrecy capacity of the \textit{non-degraded} wiretap channel was determined in \cite{WTC_IT_78}. The general formula for the wiretap channel can be found in \cite{bloch2013strong}. Although \cite{Wyner_WTC_75} and \cite{WTC_IT_78} provide the secrecy capacity for the wiretap channel, the proofs rely on Fano's inequality, and therefore, only a weak converse can be shown.

The strong converse property was first proposed in \cite{wolfowitz1957coding} for the point-to-point channel, and has received significant  attention recently due to the study of finite block-length channel coding rate \cite{Hayashi_IT_09, Yury_IT_10, Tan_now}. For the point-to-point channel, the strong converse property states that when the transmission rate is above the capacity, the asymptotic error probability goes to $1$. This implies that if we allow a potentially non-zero asymptotic error probability $\epsilon \in [0,1)$, the maximal transmission rate is still the same as the capacity, which only allows $\epsilon=0$. That is, allowing a non-zero error probability does not increase the maximal rate. Reference \cite{VH_IT_94} builds equivalent conditions for the strong converse property using the information spectrum method for the point-to-point channel, and \cite[Section~3.7]{Han_Book_03} extends it to channels with cost constraints.

The maximal transmission rate for the wiretap channel is constrained by two constraints: reliability and security. Let $\epsilon$ denote the asymptotic error probability for the reliability constraint, and let $\delta$ denote the variational distance for the secrecy constraint\footnote{There are various kinds of secrecy constraints \cite[Proposition 1]{bloch2013strong}. For instance, \cite{Wyner_WTC_75} and \cite{WTC_IT_78} use normalized mutual information, and \cite{Tan_IFS_15} and \cite{Hayashi_Allerton_14} use variational distance. We use normalized mutual information in this paper.}. Reference \cite{Tan_IFS_15} extends the method in \cite[Section~3.7]{Han_Book_03} to show the strong converse property for $(\epsilon, \delta)\in [0,1) \times \{0\}$ for the degraded wiretap channel, and name it \textit{partial strong converse} to account for the strict secrecy constraint. Reference \cite{Hayashi_Allerton_14} utilizes the relationship between the wiretap channel with feedback and secret key agreement \cite{tyagi2015converses} to show that the strong converse property holds when $\epsilon+ \delta<1$ for the degraded wiretap channel. For the degraded quantum wiretap channel, \cite{winter2016pretty} develops a ``pretty strong'' converse. Reference \cite{graves2015equal} develops strong Fano's inequalities based on image size characterization, and shows that the partial strong converse property holds for the non-degraded wiretap channel.

Recently, a new strong converse technique has been proposed in \cite{Oohama_ISIT_15, Oohama_ISIT_15_2, Oohama_Arxiv_16_ABC, Oohama_Arxiv_16_GPC, Oohama_Arxiv_16_source}. This technique is based on a novel usage of information spectrum method \cite{Han_Book_03} and a new recursive bounding method. The usage of information spectrum method provides an upper bound for the probability of correct decoding, and the recursive bounding method plays a role similar to single-letterization in the weak converse proof. It bounds the exponent function of the probability of correct decoding, and therefore shows that the probability of correct decoding goes to zero exponentially when the rate is above the capacity. This technique is general and has been applied to degraded broadcast channels in \cite{Oohama_ISIT_15}, degraded broadcast channels with feedback in \cite{Oohama_ISIT_15_2}, asymmetric broadcast channels in \cite{Oohama_Arxiv_16_ABC}, state dependent channels in \cite{Oohama_Arxiv_16_GPC}, and Wyner-Ziv coding in \cite{Oohama_Arxiv_16_source}.

Inspired by this new technique, we show that the partial strong converse property holds for the non-degraded wiretap channel. We utilize the information spectrum method and Chernoff-Cramer bound to upper bound the probability of correct decoding. Under the condition that the leakage vanishes asymptotically, we show that the exponent function of the probability of correct decoding is strictly negative when the transmission rate is higher than the secrecy capacity. Thus, the probability of correct decoding decays to zero exponentially, and the partial strong converse property holds. The main difference between our work and \cite{Oohama_ISIT_15, Oohama_ISIT_15_2, Oohama_Arxiv_16_ABC, Oohama_Arxiv_16_GPC, Oohama_Arxiv_16_source}  is that we do not construct the auxiliary distributions for the recursive bounding method for the purposes of single-letterization. Therefore, our method can be extended to channels with multi-letter characterizations \cite{dobrushin1963general, Meulen_ISIT_71, Ahlswede_ISIT_71, van1975random} for their capacity regions.

\section{Problem setting and Main Results}
\subsection{System Model and Definitions}

A wiretap channel consists of a transmitter (Alice) who wishes to send a message uniformly distributed in $\cM_n$ to a legitimate receiver (Bob) secretly in the presence of an eavesdropper (Eve) through a channel $W^n: \cX^n \rightarrow \cY^n \times \cZ^n$. $\cX$ denotes the input alphabet for Alice, while $\cY$ and $\cZ$ denote the output alphabets for Bob and Eve, respectively. $\cX$, $\cY$ and $\cZ$ are finite. Here, we consider the discrete memoryless channel, and therefore, we have
\begin{align} \label{channel}
W^n(y^n,z^n|x^n) = \prod_{t=1}^{n} W(y_t,z_t|x_t),
\end{align}
where $W(y,z|x)$ is the conditional probability mass function (pmf) of the channel.

In the following, $X^n$ denotes a random variable taking values in $\cX^n$, and the elements of $\cX^n$ are denoted by $x^n$. The pmf of random variable $X^n$ is denoted by $p_{X^n}$. Similar notation also applies to other random variables. To satisfy the secrecy constraint, we require
\begin{align} \label{secrecy_condition}
\lim_{n\rightarrow \infty}\frac{1}{n} I(M_n;Z^n)=0.
\end{align}

The encoder $\phi^{(n)}$ maps the message $m\in \cM_n$ to a codeword $x^n \in \cX^n$. We allow the encoder $\phi^{(n)}$ to be a stochastic encoder and denote it as $\phi^{(n)} = \{\phi^{(n)} (x^n| m)\}_{(m,x^n) \in \cM_n \times \cX^n }$, where $\phi^{(n)} (x^n| m)$ is a conditional pmf. The decoder is denoted by $\psi^{(n)}$ such that $\psi^{(n)}: \cY^n \rightarrow \cM_n$.
The joint pmf on $\cM_n \times \cX^n \times \cY^n \times \cZ^n$ is given by
\begin{align} \label{joint_pmf}
p_{M_nX^nY^nZ^n}&(m,x^n,y^n,z^n)      \notag  \\
&=\frac{1}{|\cM_n|}\phi^{(n)}(x^n|m) \prod_{t=1}^n W(y_t,z_t|x_t).
\end{align}

The average probability of correct decoding is given by
\begin{align}
\mP_c^{(n)}&= \mP_c^{(n)} (\phi^{(n)}, \psi^{(n)})       \\
           &\triangleq \mPr\{ \psi^{(n)}(Y^n) = M_n \}.
\end{align}
The average probability of error is $\mP_e^{(n)}=1-\mP_c^{(n)}$. For fixed $\epsilon\in[0,1)$, a rate $R$ is $\epsilon$\textit{-achievable} if there exists a sequence of codes $\{ \phi^{(n)}, \psi^{(n)}  \}_{n=1}^\infty$ such that
\begin{align} \label{limsup_Pe}
&\limsup_{n\rightarrow \infty} \mP_e^{(n)} \leq \epsilon,   \\
&\liminf_{n\rightarrow \infty} \frac{1}{n}\log |\cM_n| \geq R.
\end{align}

In this work, we consider the partial strong converse property. Therefore, we require the code to satisfy \eqref{secrecy_condition}.
Let $\cC_s(\epsilon|W)$ denote the maximal $\epsilon$\textit{-achievable} rate satisfying \eqref{secrecy_condition} through the wiretap channel $W(y,z|x)$. From \cite{WTC_IT_78}, we have
\begin{align} \label{secrecy_capacity}
\cC_s(0|W)=\max_{p\in\cP(W)} I(U;Y)-I(U;Z),
\end{align}
where $\cP(W)$ is defined as
\begin{align}
\cP(W) \triangleq \bigg\{  p_{UXYZ}&(u,x,y,z): |\cU|\leq |\cX|,   \notag \\
                          & p_{YZ|X}(y,z|x)=W(y,z|x), \notag \\
                          & U \rightarrow X \rightarrow (Y,Z)  \bigg\}.
\end{align}

\subsection{Main Result}
\begin{theorem} \label{thm_1}
For a discrete memoryless non-degraded wiretap channel $W(y,z|x)$, the partial strong converse property holds, i.e., for $\epsilon\in[0,1)$, for any code $(\phi^{(n)},\psi^{(n)})$ satisfying \eqref{secrecy_condition},
\begin{align}
\cC_s(\epsilon|W)=\cC_s(0|W).
\end{align}
\end{theorem}

The proof is provided in Section~\ref{Sec_proof}. It is inspired by the new strong converse proof technique developed in \cite{Oohama_ISIT_15, Oohama_ISIT_15_2, Oohama_Arxiv_16_ABC, Oohama_Arxiv_16_GPC, Oohama_Arxiv_16_source}. We first use Verdu-Han bound and Chernoff-Cramer bound to upper bound the probability of correct decoding. Then, we focus on bounding the exponent function. We show that when the transmission rate is above  $\cC_s(0|W)$ and \eqref{secrecy_condition} is satisfied, the probability of correct decoding decays to zero exponentially fast, and the partial strong converse property holds.

\section{Proof of the Main Result} \label{Sec_proof}

Consider a sequence of codes for which \eqref{secrecy_condition} is satisfied. Therefore, for each $\delta>0$, there exists $n_0$ such that $\forall n > n_0$, we have
\begin{align} \label{secrecy_condition_delta}
	\frac{1}{n} I(M_n;Z^n) < \delta.
\end{align}
In the following, we consider $n>n_0$.

\begin{lemma}(Verdu-Han \cite[Theorem 4]{VH_IT_94})  \label{Lemma_1}
For any $\eta>0$ and for any $(\phi^{(n)},\psi^{(n)})$ satisfying $\frac{1}{n} \log |\cM_n| \geq R$, we have
\begin{align}
\mP_c^{(n)}(\phi^{(n)},&\psi^{(n)}) \leq p_{M_nX^nY^nZ^n} \bigg\{        \notag \\
 R \leq&  \frac{1}{n} \log \frac{p_{Y^n|M_n}(Y^n|M_n)}{p_{Y^n} (Y^n)}+\eta \bigg\} + e^{-n\eta}. \label{L1_1}
\end{align}
\end{lemma}

\begin{lemma}(Chernoff-Cramer) \label{Lemma_2}
For any real valued random variable $A$ and any $\theta>0$, we have
\begin{align} \label{CC_inequality}
\Pr \{ A \geq a \} \leq \exp \{ -[ \theta a - \log \mE[\exp(\theta A)]    ]  \}.
\end{align}
\end{lemma}

By Lemmas~\ref{Lemma_1} and \ref{Lemma_2}, we have the following lemma.
\begin{lemma} \label{Lemma_3}
For any $\theta>0$, and for any $(\phi^{(n)},\psi^{(n)})$ satisfying $\frac{1}{n} \log |\cM_n| \geq R$, we have
\begin{align} \label{L3_eq}
\mP_c^{(n)}&(\phi^{(n)},\psi^{(n)}) \notag \\
           & \leq \exp\Big\{n \big[ \theta \eta -\theta R  + \frac{1}{n} \COmega  \big] \Big\} \notag \\
           &\quad  + e^{-n\eta}
\end{align}
where $\COmega$ is defined as
\begin{align} \label{def_COmega}
&\COmega \notag \\
&\qquad \quad \triangleq \log \mE_{p^{(n)}} \Bigg[ \bigg\{ \frac{p_{Y^n|M_n}(Y^n|M_n)}{p_{Y^n}(Y^n)}  \bigg\}^\theta  \Bigg]
\end{align}
where $p^{(n)}=p_{M_nX^nY^nZ^n}$ is defined in \eqref{joint_pmf}.
\end{lemma}

\begin{Proof}
We define the random variable $B$ as
\begin{align}
B\triangleq \frac{1}{n} \log \frac{p_{Y^n|M_n}(Y^n|M_n)}{{p_{Y^n}} (Y^n)}-R.
\end{align}
Then, by \eqref{L1_1} in Lemma~\ref{Lemma_1}, we have
\begin{align}
\mP_c^{(n)}&(\phi^{(n)},\psi^{(n)})  \notag \\
&\leq p_{M_nX^nY^nZ^n} \{ B\geq-\eta \} + e^{-n\eta} \\
& =   p_{M_nX^nY^nZ^n} \{ nB\geq-n\eta \} + e^{-n\eta}
\end{align}
By identifying $A=nB$, $a=-n\eta $ and applying \eqref{CC_inequality} in Lemma~\ref{Lemma_2}, we have
\begin{align}
&\mP_c^{(n)}(\phi^{(n)},\psi^{(n)})                                                \notag \\
&\quad\leq \exp\{-[ \theta(-n \eta )- \log \mE_{p^{(n)}} [\exp(n\theta B)]  ] \} + e^{-n\eta} \\
&\quad=    \exp\Big\{n \big[ \theta \eta -\theta R  + \frac{1}{n} \COmega  \big] \Big\} \notag \\
&\quad \quad + e^{-n\eta}.
\end{align}
\end{Proof}

Let $\cP^{(n)}(W)$ be a set of all pmfs $p_{M_nX^nY^nZ^n}$ on $\cM_n \times \cX^n \times \cY^n \times \cZ^n$ defined in \eqref{joint_pmf}. Moreover, define the subset of all pmfs in $\cP^{(n)}(W)$ that satisfy the secrecy constraint in (\ref{secrecy_condition_delta}) as,
\begin{align} \label{P_n_delta_def}
\cP^{(n)}_\delta(W) = \bigg\{ p_{M_nX^nY^nZ^n}:\quad  & p_{M_nX^nY^nZ^n} \in  \cP^{(n)}(W),        \notag \\
                                & \frac{1}{n} I(M_n;Z^n) < \delta    \bigg\}.
\end{align}

In order to bound the exponent function in \eqref{L3_eq}, we define the communication potential $\BComega$ as follows:
\begin{align} \label{communication_potential}
\BComega \triangleq \max_{p^{(n)} \in \cP^{(n)}_\delta(W) }  \frac{1}{n} \COmega.
\end{align}

We note that we do not construct the auxiliary distributions for single-letterization. Therefore, the definition of communication potential is different from the definitions given in \cite{Oohama_ISIT_15, Oohama_ISIT_15_2, Oohama_Arxiv_16_ABC, Oohama_Arxiv_16_GPC, Oohama_Arxiv_16_source}. References \cite{Oohama_ISIT_15, Oohama_ISIT_15_2, Oohama_Arxiv_16_ABC, Oohama_Arxiv_16_GPC, Oohama_Arxiv_16_source} apply a new technique called recursive bounding method for the single-letterized exponent. Here, we keep the multi-letterized form, and connect it to the proof of the weak converse.

From \eqref{L3_eq} in Lemma~\ref{Lemma_3}, we have
\begin{align} \label{L3_eq_ext}
\mP_c^{(n)}(\phi^{(n)},\psi^{(n)}) & \leq \exp\Big\{n \big[ \theta \eta -\theta R  + \BComega  \big] \Big\} \notag \\
&\quad  + e^{-n\eta}.
\end{align}

We remark that there are two main factors affecting the exponent function in \eqref{L3_eq_ext}: the code rate $R$ and the communication potential $\BComega$.

To further study \eqref{L3_eq_ext}, we show that $\COmega$ has the properties listed in Lemma~\ref{Lemma_4} below. Lemma~\ref{Lemma_4} connects communication potential to the mutual information expression. We note that the properties listed in Lemma~\ref{Lemma_4} are first obtained in \cite{Oohama_ISIT_15, Oohama_ISIT_15_2, Oohama_Arxiv_16_ABC, Oohama_Arxiv_16_GPC, Oohama_Arxiv_16_source}.

\begin{lemma} \label{Lemma_4} \leavevmode
\begin{enumerate}
\item $\COmega$ is a convex function of $\theta>0$, where  $\COmega$ is defined in \eqref{def_COmega}.
\item
\begin{align}  \label{L4_eq_1}
&\lim_{\theta \rightarrow 0^+} \frac{ \frac{1}{n} \COmega}{\theta}=\frac{1}{n} I(M_n;Y^n)
\end{align}
\item
For each $\Delta>0$, there exists $\theta>0$ such that
\begin{align} \label{L4_eq}
\BComega \leq \theta \left[\max_{p^{(n)} \in \cP_{\delta}^{(n)} } \frac{1}{n} I(M_n; Y^n) + \frac{\Delta}{2} \right],
\end{align}
where $\BComega$ is given in \eqref{communication_potential}.
\end{enumerate}		
\end{lemma}

\begin{Proof}
To simplify the notation, define
\begin{align}
\ua &\triangleq (m,y^n), \\
\uA &\triangleq (M_n,Y^n), \\
\ucA &\triangleq \cM_n \times \cY^n, \\
\rho(\ua) &\triangleq  \log  \frac{p_{Y^n|M_n}(y^n|m)}{p_{Y^n}(y^n)}, \\
\xi(\theta) &\triangleq \COmega.
\end{align}

From \eqref{def_COmega}, we have
\begin{align} \label{convexity}
\xi(\theta) = \COmega = \log \left[  \sum_{\ua\in\ucA} p^{(n)}_{\uA}(\ua)e^{\theta\rho(\ua)} \right].
\end{align}
We evaluate the second derivative to show the convexity of $\theta$
\begin{align}
&\xi'(\theta)=e^{-\xi(\theta)}\left[ \sum_{\ua\in\ucA} p^{(n)}_{\uA}(\ua) \rho(\ua) e^{\theta\rho(\ua)} \right],  \label{1st_derivative}   \\
&\xi''(\theta)=e^{-2\xi(\theta)}  \notag \\
& \quad \times \left[ \sum_{\ua,\ub\in\ucA} p^{(n)}_{\uA}(\ua) p^{(n)}_{\uA}(\ub) \frac{ \{ \rho(\ua)-\rho(\ub)  \}^2  }{2} e^{\theta\{\rho(\ua)+\rho(\ub)  \}}\right].  \label{2nd_derivative}
\end{align}
From \eqref{2nd_derivative}, since $\xi''(\theta)\geq0$, $\COmega$ is a convex function of $\theta>0$. Hence, part 1) holds.

For part 2), consider $\theta=0$ in \eqref{1st_derivative}, and $\xi(0)=0$. We have
\begin{align} \label{eq_part2}
\xi'(0) = I(M_n;Y^n).
\end{align}

For part 3), given $\Delta>0$, define the following function:
\begin{align} \label{zeta}
\zeta(\theta) \triangleq \BComega - \theta\left[\max_{p^{(n)} \in \cP^{(n)}_\delta(W) } \frac{1}{n} I(M_n; Y^n) + \frac{\Delta}{2} \right].
\end{align}
The function $\zeta(\theta)$ has the following properties:
\begin{align}
&\zeta(0)=0,   \label{L4_1}  \\
&\zeta'(0)=-\frac{\Delta}{2} < 0,  \label{L4_2} \\
&\zeta''(\theta)= \frac{1}{n} \xi''(\theta) \geq 0.   \label{p2}
\end{align}
By the definition given in \eqref{communication_potential} and \eqref{L4_eq_1}, \eqref{L4_2} holds. \eqref{p2} follows \eqref{2nd_derivative}. For each $\Delta>0$, there exists $f(\Delta)>0$ such that for $\theta\in(0,f(\Delta)]$, we have $\zeta(\theta) \leq 0$. Finally, \eqref{L4_eq} holds. 	
\end{Proof}

Now, consider a code with rate
\begin{align}
R=\cC_s(0|W) + 4 \delta.
\end{align}
We can further bound $\BComega$ in \eqref{L4_eq} as in the following lemma.

\begin{lemma} \label{Lemma_5}
For each $\delta>0$, there exists $\theta>0$ such that
\begin{align} \label{L5_eq}
\BComega \leq \theta \left[ \cC_s(0|W) + 3 \delta \right],
\end{align}
where $\BComega$ is given in \eqref{communication_potential}.		
\end{lemma}
\begin{Proof}
Take $\Delta=4\delta$. For the fixed $\Delta>0$, from \eqref{L4_eq} in Lemma~\ref{Lemma_4}, there exits a $\theta>0$ such that
\begin{align}
&\BComega  \leq \theta \left[\max_{p^{(n)} \in \cP_{\delta}^{(n)} } \frac{1}{n} I(M_n; Y^n) + 2\delta \right]   \\
& \leq \theta \left[\max_{p^{(n)} \in \cP^{(n)} } \frac{1}{n} I(M_n; Y^n) + \left(\delta - \frac{1}{n} I(M_n; Z^n)  \right) + 2\delta \right]  \label{L5_a} \\
& \leq \theta \left[ \cC_s(0|W)  + 3 \delta \right] ,  \label{L5_b}
\end{align}
where \eqref{L5_a} holds due to the definition in \eqref{P_n_delta_def}, and \eqref{L5_b} holds due to \cite{WTC_IT_78}\cite[Section 22.1.2]{Gamal_book}.
\end{Proof}

We are now ready to prove Theorem~\ref{thm_1}. We consider a sequence of codes for which \eqref{secrecy_condition} is satisfied. Therefore, given $\delta>0$, there exists $n_0$ such that $\forall n > n_0$, we have \eqref{secrecy_condition_delta}.

Now, assume that the code rate, $R$, equals to $\cC_s(0|W)+4\delta$. From Lemma~\ref{Lemma_3}, we upper bound the probability of correct decoding as \eqref{L3_eq}. We further upper bound \eqref{L3_eq} as \eqref{L3_eq_ext}. Now, taking $\Delta=4\delta$, from 3) in Lemma~\ref{Lemma_4}, there exists $\theta>0$ such that \eqref{L4_eq} holds, which can be further upper bounded by \eqref{L5_eq} in Lemma~\ref{Lemma_5}. Therefore, \eqref{L3_eq_ext} becomes
\begin{align}
\mP_c^{(n)}(\phi^{(n)},\psi^{(n)}) & \leq \exp\Big\{n \big[ \theta \eta -\theta R  + \theta \left[ \cC_s(0|W) + 3 \delta \right]  \big] \Big\} \notag \\
&\quad  + e^{-n\eta} \\
&=\exp\Big\{n \theta \big[  \eta   - \delta   \big] \Big\}+ e^{-n\eta}.
\end{align}

Finally, by picking $\eta=\frac{\delta}{2}$, we have
\begin{align}
\mP_c^{(n)}(\phi^{(n)},\psi^{(n)})  \leq e^{-n \theta  \frac{\delta}{2}    }+ e^{-n\frac{\delta}{2}}.
\end{align}
Thus, the probability of correct decoding decays to zero exponentially fast. For each $\delta>0$ and $R \geq \cC_s(0|W)+4\delta$, we have shown $\limsup_{n \rightarrow \infty} \mP_e^{(n)}=1$. Therefore, for fixed $\epsilon \in [0,1)$, if \eqref{secrecy_condition} and \eqref{limsup_Pe} are satisfied, then $R<\cC_s(0|W)+4\delta$. Hence, we have $\cC_s(\epsilon|W)=\cC_s(0|W)$, completing the proof of Theorem~\ref{thm_1}.

\section{Conclusions and remarks}
In this work, we proved the partial strong converse property for the discrete memoryless non-degraded wiretap channel, for which we require the leakage to the eavesdropper to vanish. Our work is based on the information spectrum method and Chernoff-Cramer bound for upper bounding the probability of correct decoding. We focus on bounding the exponent function of the probability of correct decoding. There are two main factors affecting the exponent function: the transmission rate and the communication potential. When the transmission rate is higher than the secrecy capacity, we show that the exponent is strictly negative. Therefore, the probability of correct decoding decays to zero exponentially fast, which implies that the partial strong converse property holds.

We remark that the proof of Theorem~\ref{thm_1} can also be adapted to the proof of the strong converse for the point-to-point discrete memoryless channel. First, we define $\cC(0|W)$ as the channel capacity for the point-to-point channel. For this channel, $p^{(n)}=p_{M_nX^nY^n}$ in Lemma~\ref{Lemma_3}. Since there is no security constraint, the communication potential for the point-to-point channel becomes
\begin{align}
\BComega \triangleq \max_{p^{(n)} \in \cP^{(n)}(W) }  \frac{1}{n} \COmegaPtoP.
\end{align}
Moreover, \eqref{L4_eq} in Lemma~\ref{Lemma_4} becomes
\begin{align}
\BComega \leq \theta \left[\max_{p^{(n)} \in \cP^{(n)} } \frac{1}{n} I(M_n; Y^n) + \frac{\Delta}{2} \right].
\end{align}
Therefore, given $\Delta>0$, for a code with $R \geq \cC(0|W)+\Delta$, we have $\limsup_{n \rightarrow \infty} \mP_e^{(n)}=1$.

We also remark that the proof of Theorem~\ref{thm_1} can also be adapted to the proof of the strong converse for the discrete memoryless multiple access channel, discrete memoryless interference channel and discrete memoryless broadcast channel. Although we do not know the single-letterized capacity region expressions for interference and broadcast channels, we have the multi-letterized capacity region expressions \cite{Ahlswede_ISIT_71, van1975random}. With the multi-letterized capacity region expressions \cite{Meulen_ISIT_71, Ahlswede_ISIT_71, van1975random}, we can show that these channels satisfy the strong converse property. For the multiple access channel, the strong converse property has been reported in \cite{dueck1981strong, ahlswede1982elementary}. Our method can be viewed as an alternative proof without resorting to the wringing technique. For broadcast channels, the strong converse properties reported so far are for those with known single-letterized capacity regions, such as the degraded broadcast channel \cite{ahlswede1976bounds, Oohama_ISIT_15} and the broadcast channel with degraded message sets \cite{korner1977general, Oohama_Arxiv_16_ABC}.

\bibliographystyle{unsrt}
\bibliography{IEEEabrv,reference_jimtm}

\end{document}